\documentstyle[12pt]{article}

\newcommand{\beq}{\begin{equation}}
\newcommand{\eeq}{\end{equation}}

\begin{document}
\def\lag{\langle}
\def\rag{\rangle}
\title{  RENORMALISATION GROUP AIDED\\
FINITE TEMPERATURE REDUCTION\\
 OF QUANTUM FIELD THEORIES\\
}
\author{{A. Patk\'os$^{1}$, P. Petreczky$^{2}$ and J. Polonyi$^{1,3}$}\\
}
\vfill
\footnotetext[1]{{\em Dept. of Atomic Physics, E\"otv\"os University,
H-1088, Puskin u. 5-7., Budapest Hungary}}
\footnotetext[2]{{\em Kossuth University,
 H-4026 Bem J. t\'er 18., Debrecen Hungary}}
\footnotetext[3]{{\em Lab. Physique Th\'eorique, Univ. L. Pasteur,
 3 rue de l'Universit\'e, 67084 Strasbourg Cedex, France}}
\maketitle
\begin{abstract}
Dimensional reduction of finite temperature quantum field
theories can be improved with help of continous renormalisation group
steps. The method is applied to the integration of the lowest non-static
($n=\pm 1$) modes of the finite temperature $\Phi^4$-theory. A second,
physically important application is the integration of the Debye-screened
${\bf A_0}(x)$ static scalar potential in the gauged SU(2) Higgs model.
\end{abstract}

\newpage

\section{ Introduction}

Thermal field theories at finite temperature can be equivalently formulated
with help of countably infinite number of three-dimensional fields representing
each quantum field (Matsubara decomposition). In the three dimensional
formulation each set of representatives can be classified at tree level
as belonging to
the group of massive ($m\sim~T$) or massless (static) components. (Fermions
are represented exclusively by massive modes.) The reduction program consists
of
integrating out the massive components, their role being reduced in this way to
influence (sometimes essentially) the interaction between static modes. The
fluctuations of these latter are widely accepted to govern temperature
driven phase transitions in quantum field theory.

The usual procedure of perturbative reduction \cite{appelquist,landsman}
is of restricted validity. Especially detailed quantitative non-universal
characteristics ($T_c$, order parameter discontinuity, surface tension, etc.)
might be very sensitive to the correct dependence of the couplings of the
reduced theory on the temperature and on the parameters of the original ($T=0$)
theory. For this one should achieve the most faithful mapping of the full
finite temperature system onto the effective theory. In case of the QCD
deconfining transition an answer has been attempted to this problem by
matching numerically
several observables of the full finite temperature theory to the values
simulated with the three-dimensional effective theory
\cite{reisz,karka,braaten,karsch}.

The use of an "exact" Renormalisation Group transformation for the local part
of the action has been practised in critical phenomena already
for two decades for the determination
of the RG-flow and of the fixed point effective potential
\cite{nicoll,wegner,hasenfratz,margaritis,tetradis,senben,morris}.
To our knowledge, however, the investigation of its usefulness
 from the point of view of the dimensional reduction appears for
the first time in the present paper.

We shall test its applicability by {\it completing the reduction}
in models where,
after a partial (perturbatively performed) reduction only finite number of
three-dimensional fields are left, but for some reason still a well-defined gap
persists between their respective thermal masses. Intuitively, perturbative
integration should hold for the most massive (highest) Matsubara modes, but
considerable improvement is expected to result from a more accurate
non-perturbative integration over modes, lying the closest to the lightest
ones.

The first question we should like to clarify is to what extent it is possible
to take into account higher loop effects in the reduction, which reflect the
interaction between the non-static modes. Our proposal is to integrate out
perturbatively only the modes with $|n|<n_0$, and proceed to a second stage of
integration with the couplings already modified in the first step. Actually,
the
application of the RG-approach in the second stage allows the realisation of a
continous feedback of the modified couplings on the process of lowering the
momentum scale for the fields to be integrated out. As a testing ground for the
procedure outlined above, we shall investigate the finite temperature
1-component $\Phi^4$ theory with the choice $n_0=1$.

A similar situation is arrived at in the gauged SU(2) Higgs model,
after having integrated
out perturbatively {\it all} non-static modes. The phenomenon of
Debye-screening
creates a new distinct mass scale $(m\sim gT)$, separating the ${\bf A_0(x)}$
static modes from the rest \cite{buchmuller,zwirner,kajantie1}.
It is then natural to continue with a
second (hierarchical) integration of the 3 ${\bf A_0}$-field components,
and apply MC-simulation methods to the
resulting three-dimensional gauge+Higgs effective system.
Propositions for perturbative realisation of this program have been already
published \cite{jako1,farakos1}.

In this paper we shall expose and test the application of the "exact"
RG-transformation to situations corresponding to the two problems shortly
outlined above. We are going to use a local version of this transformation,
therefore effects of the field renormalisation will not be considered in the
present paper.

In Section 2 the general strategy of the application of Renormalisation Group
motivated reduction will be discussed in detail. Thorough
qualitative description of some intrinsic features of the procedure will be
given, providing insight into the temperature range where the proposed RG aided
reduction is expected to work. In Section 3 we introduce our test-models
mentioned above (the lengthy, but interesting excercise of partial perturbative
reduction of the $\Phi^4$-theory to the $n_0=1$ level is relegated to the
Appendix). In Section 4 the RG aided reduction will be realised on these
systems
under the
assumption that the local potential density of the resulting effective theory
can be well truncated to the linear combination of a
finite number of lowest dimensional operators. The
arising finite coupled set of differential infrared RG-flow
equations will be carefully
integrated, and the resulting couplings compared with the characteristics of
the
fully perturbative (sometimes higher-loop) reductions.
The summary of our results with a detailed discussion of the qualitative
features of the final light theory is presented in Section 5.

\section{The Strategy of the Renormalisation Group Aided Reduction}

The generic situation one deals with is a three-dimensional theory (arrived at
in previous -- possibly perturbative -- reduction steps) with one light
$(\phi)$
and one heavy $(\Phi)$ field in interaction. At momentum scale $\Lambda$ their
interaction potential is known
\begin{equation}
U_{\Lambda}(\Phi ,\phi )=V(\Phi ,\phi ,\Lambda ,T)
\end{equation}
and in particular one has an explicit expression for the corresponding
mass-terms:
\begin{eqnarray}
&
{\partial^2V\over\partial\phi^2}_{|\Phi =\phi
=0}=m^2+a_{\phi}T^2-b_{\phi}\Lambda T\equiv M^2_{\phi}(\Lambda ,T),\nonumber\\
&
{\partial^2V\over \partial \Phi^2}_{|\Phi =\phi
=0}=m^2+a_{\Phi}T^2-b_{\Phi}\Lambda T\equiv M^2_{\Phi}(\Lambda ,T).
\end{eqnarray}
These masses are renormalised from the four-dimensional ($T=0$) point of view,
but contain explicitly the induced three-dimensional "counterterm"-contribution
depending linearly on the cut-off. If the reduction is realised with higher
loop accuracy the linear cut-off dependence might be modulated by logarithmic
terms.

By assumption there is a relation between the values of the two bare masses at
 high momentum:
\begin{equation}
M^2_{\phi}(\Lambda ,T)<M^2_{\Phi}(\Lambda ,T).
\end{equation}
($m^2$ is the mass parameter of the original theory renormalised by a
condition imposed at $T=0$.)

The integration of the heavy field $\Phi$ will be performed in infinitesimal
steps, integrating in each step over its Fourier components lying in the thin
layer $k\leq |{\bf p}|\leq k+\Delta k\equiv K, \Delta k\rightarrow 0$, and
feeding back the resulting modified couplings into the next integration step.
For an infinitesimally thin layer the 1-loop contribution will be dominant. It
has to be evaluated in a background of the fields $\phi_L({\bf x}), \Phi_L({\bf
x})$ composed exclusively of Fourier components of momenta $|{\bf p}|<k$. For
the
computation of the potential energy density $\phi_L, \Phi_L$ can be set
constant.

The matrix characterising the high spatial frequency fluctuations $\Phi_H({\bf
x)}$ at the Gaussian (1-loop) level is given by
\begin{eqnarray}
&
U_K''(\Phi_L,\phi_L)\equiv{\partial^2U_K\over
\partial\Phi_p\partial\Phi_{-p}}_{|p=K}(\Phi_L,\phi_L)=\nonumber\\
&
=M^2_{\Phi}(K,T)+c_1(K,T)\Phi^2_L+c_2(K,T)\phi^2_L+{\cal
O}(\Phi^4_L,\Phi^2_L\phi^2_L,\phi^4_L)
\end{eqnarray}
The change in the potential energy is governed by the familiar differential
equation
\begin{equation}
k{\partial U_k\over \partial k}=-{1\over 4\pi^2}k^3\ln
(k^2+U_k''(\Phi_L,\phi_L)).
\label{logrg}
\end{equation}
This equation should be integrated "downwards" on the momentum scale $k$, with
the initial conditions (1),(2) fixed at $k=\Lambda$.

For small field values of $\Phi_L,\phi_L$ the quadratic form (4) is well
approximated by the terms written out explicitly. Near the cut-off $\Lambda$
the large negative contribution to $M^2_{\Phi}(\Lambda ,T)$, linear in
$\Lambda$ is compensated by adding in the argument of the logarithm to it
$k^2=\Lambda^2$ and the contribution to $U_k$ is
real. In case $M^2_{\Phi}(k,T)$ would not run appropriately with $k$, the
argument of the logarithm on the right hand side of (5) would change sign at a
certain scale $k_{\Phi}(T)$, which would mean that a consistent integration of
$\Phi$ at that temperature would prove impossible.

Since $M^2_{\Phi}(k=0,T)$ is the
renormalised mass of the heavy field, for sufficiently high temperature one can
expect with certainty it to be positive. Then the reduction can be performed
meaningfully. There is a threshold temperature $T_{\Phi}$ at which $M^2_{\Phi}$
crosses zero first as the temperature is being lowered. It presents a limit
in temperature to the consistent reduction with the present simple formulation
of the Renormalisation Group idea. It could well be that by
appropriate fine-tuning
 of the cut-off procedure one could avoid the reduced potential
becoming imaginary, in the same way as it has been demonstrated for the
effective potential by \cite{tetradis,ringwald}. However, this
would make the actual procedure rather complicated and it seems to us not
necessary for the purposes of the reduction on the basis
of the following simple argument.

We would like to study the immediate neighbourhood of the transition
temperature $T_c$. Therefore the reduction procedure is useful only if one
finds $T_c>T_{\Phi}$. It is plausible that the heavy-light mass-inequality (3)
will persists all the way when rolling down on the momentum scale $k$:
\begin{equation}
M^2_{\phi}(k,T)<M^2_{\Phi}(k,T).
\end{equation}
Then repeating the above logical steps for the light degree of freedom we are
led to the conclusion that the instability of the constant background field
$\phi_L$, signalling the phase transition, will indeed occur at
higher temperature, than $T_{\Phi}$.

\section{Finite temperature effective models with heavy-light
field hierarchy}

The challenge we face is to find the simplest representation of the
{\it electroweak theory} at finite temperature containing the smallest number
of
degrees of freedom, without loosing any essential aspect
of the phase transition.
The perturbative integration over $n\neq 0$ Matsubara modes seems to be
intuitevely
correct since their mass scale $2\pi T$ is much larger than the scale
characterising nonperturbative quantities, like magnetic screening. This
latter is characterized by the scale $g^2T$ \cite{rebhan}, and
for small $g$ it is even smaller than the "electric"
Debye-screening mass-scale $gT$.

The effective three-dimensional theory has been derived on the 1-loop level
in different renormalisation schemes by different groups \cite{farakos1,jako2}.
Following \cite{jako2} we use the form obtained with momentum cut-off
regularisation in the scheme where the location and the value of the
non-trivial
minimum of the potential energy density of the effective model is kept at the
values fixed on the classical level \cite{linde}. Its expression is the
following:
\begin{equation}
S[\bar A_i,\bar A_0, \phi] = \int d^3x \bigl[L_{\rm kin}+U(\bar A_0,
\phi)\bigr],
\end{equation}
where
\begin{eqnarray}
&
L_{\rm kin}={1\over 4}\bar F_{ij}^{a}\bar F_{ij}^{a}+{1\over
2}[(\partial_{i}+i\bar g\bar A_{i})\phi ]^{+}[(\partial_{i} +i\bar
g\bar A_{i})\phi ]+{1\over 2}(\partial_{i}\bar A_{0}+\bar
g\epsilon^{abc}\bar A_{i}^{b}\bar A_{0}^{c})^{2},\nonumber\\
&
U_{{\rm dim} 4}(\bar A_0,\phi)={1\over 2}m_{\phi}^{2}\phi^{+}\phi+
{1\over 2}m_{D}^{2}\bar A_{0}^{2}+{1\over 8}\bar
g^{2}\bar A_{0}^{2}\phi^{+}\phi+{T\hat\lambda\over
24}(\phi^{+}\phi)^{2}\nonumber\\
&
{}~~~~~+{17\beta\bar g^{4}\over
192\pi^{2}}(\bar A_{0}^{2})^{2}-[{5{\bar g}^2\over 4\pi^2}A_0^2
+({9{\bar g}^2\over 4}+\bar\lambda){1\over 4\pi^2}\phi^{+}\phi ]\Lambda,
\end{eqnarray}
\begin{equation}
m_{\phi}^{2}=\hat m^{2}+({3\over 16}\bar g^{2}+{\bar\lambda\over
12})T,~~~~m_{D}^{2}={5\over 6}\bar g^{2}T+{m^{2}\bar g^{2}\over
8\pi^{2}T},
\label{ewdefmass}
\end{equation}
\begin{eqnarray}
&
\hat m^{2}=m^{2}
(1-{1\over 32\pi^{2}}\bigl(({9\over 2}g^{2}+\lambda)\ln{3g^{2}v_{0}^{2}
\over 4T^{2}}+\lambda\ln{\lambda v_{0}^{2}\over  3T^{2}}\bigr)
-{1\over 128\pi^{2}}(45g^{2}+20\lambda+{27g^{4}\over \lambda})),\nonumber\\
&
\hat\lambda =\lambda -{3\over 8\pi^{2}}(g^{4}({3\over 8}\ln {g^{2}v_{0}^{2}
\over 4T^{2}}-{3\over 2}\ln{g^{2}v_{0}^{2}\over \sqrt{2}T^{2}})+
{\lambda^{2}\over 4}\ln{\lambda v_{0}^{2}\over 3T^{2}}+3({3g^{2}\over 4}+{
\lambda\over 6})^{2}\ln{3g^{2}v_{0}^{2}\over 4T^{2}})-\nonumber\\
&
{}~~~~~~~~~~~~~~~~~~-{9\over 16\pi^{2}}({9g^{4}\over 16}+{3g^{2}\lambda\over
4}+{\lambda^{2}\over 3}).
\end{eqnarray}

Perturbative investigations of the phase transition have shown that near the
transition point the screening mass of the $A_0$ $(\sim gT$)
quanta is considerably larger than the masses of the Higgs and of the magnetic
gauge field fluctuations. For this reason we assume
(and shall test subsequently the consistency of the assumption) that
in this model the
$A_0$ multiplet represents the heavy fields which can be
integrated out in some careful procedure.

 An analogous situation can be created by construction also within the familiar
1-component $\Phi^4$-model, if instead of integrating over all non-static
fields at once, as a first step only the $|n|\geq 2$ modes are integrated out.
As a result an effective theory with a light ($M\sim \lambda T$) and a heavy
($M\sim 2\pi T$) field arises. (One should note the difference in the size of
the heavy and light mass-scales between the two models).

We expect the subsequent integration of the
heavy field will display also some features of the 2-loop reduction, but
not any really dramatic deviation from the standard 1-step reduction. Applying
a non-perturbative integration technique to the heavy component one can test
the stability of the standard reduction (in the weak coupling regime)
against this expectation.

The derivation of the effective heavy-light theory in the $\Phi^4$-theory
 represents a technically quite involved {\it perturbative} procedure. Since
in the main text we wish to concentrate on features of the non-perturbative
reduction, we describe  details of the perturbative part of the reduction
in Appendix A. The result of
this  (partial) reduction is given by the following effective action:
\begin{eqnarray}
&
S_{3D}[\phi_0, \phi_1 ]=\int d^3x\bigl\{{1\over 2}(\partial_i\phi_0)^2+
|\partial_i\phi_1|^2+{1\over 2}(\phi_0^2+2|\phi_1|^2)(m^2+{13\bar\lambda T\over
24}- {3\bar\lambda\Lambda\over 4\pi^2})\nonumber\\
&
{}~~~~~~~~+(2\pi T)^2|\phi_1|^2+{\bar\lambda\over
24}[\phi_0^4+12(1+a\lambda)\phi_0^2|\phi_1|^2
+6(1+b\lambda)|\phi_1|^4]\nonumber\\
&
+U_6[\phi_0,\phi_1]\bigr\},
\end{eqnarray}
where
\begin{equation}
\phi_0=\sqrt{\beta}\Phi_0,~~~\phi_1=\sqrt{\beta}\Phi_1,~~~\bar\lambda=
\lambda T,
\end{equation}
and the numerical constants (see Appendix A)
\begin{equation}
a=3.56\times 10^{-3},~~~~~b=3.17\times 10^{-3}.
\end{equation}
The normalisation condition required at this level to ensure the 4-dimensional
ultraviolet finiteness of the theory is of the following form:
\begin{equation}
{\partial^4U_{3D}\over\partial\phi_0^4}|_{\phi_1=0}=\bar\lambda.
\end{equation}
This condition explains the presence of  finite (${\cal O}(\lambda^2)$)
corrections
to the terms $|\phi_1|^4$ and $|\phi_1|^2\phi_0^2$ above.

Also the derivation of the sixth-power contribution to the potential is
given in Appendix A. It is of the form:
\begin{equation}
\Delta U_6=\Delta U_6^{(1)}+\Delta U_6^{(2)}
\end{equation}
with the following explicit expressions for the terms on the right hand side:
\begin{eqnarray}
&
\Delta U_6^{(1)}=-T^2\lambda^2\bigl ({1\over 4}\phi_0^2|\phi_1|^4{1\over
m^2+\omega_2^2}+{1\over 36}|\phi_1|^6{1\over m^2+\omega_3^2}\bigr ),\nonumber\\
&
\Delta U_6^{(2)}=5.40183~10^{-6}
M^6+2.7566~10^{-5}\lambda^2\phi_0^2|\phi_1|^2M^2\nonumber\\
&
+6.6586~10^{-6}\lambda^2|\phi_1|^4M^2-1.6991~10^{-5}\lambda^3\phi_0^2|\phi_1|^4
\end{eqnarray}
($M^2=m^2+\bar\lambda(\phi_0^2+2|\phi_1|^2)/2, \omega_n=2\pi nT).$

This contribution is usually neglected, since it is suppressed at high
temperature by $\Phi^2/T^{2}$  relative to the quartic terms, when $|\Phi|<<T$.
It might prove interesting to include it into the non-perturbative treatment
of the $\phi_1$-integration, at least to check the validity of the
truncation at the quartic term.

In the next section the evolution equations for the potential
arising  from the non-perturbative
 integration over subsequent momentum layers of the $\phi_1$ field
will be worked out explicitly. In the approximation,
when the $\phi_0$-potential of the resulting theory is projected onto a
quartic polynomial, these equations can be exactly mapped onto the
flow equations  of the Higgs-potential of the electroweak theory induced by
 the $A_0$-integration. This allows a uniform formal
treatment of the two physically different heavy-light systems from the
point of view of the reduction.

\section{Non-perturbative integration of heavy fields (finite polynomial
approximation)}

The detailed analysis of the non-perturbative integration will be illustrated
with the $\phi_0 -\phi_1$ system. At the end of the section we describe the
corresponding results for the effective theory of the electroweak phase
transition.

The first step of the procedure consists of the application of two subsequent
Hubbard-Stratonovich transformations, through which the operators
$\sim |\phi_1|^6$ and $\sim |\phi_1|^4$ can be expressed in a form apparently
quadratic in $\phi_1$ with help of two auxiliary Gaussian
variables $(\chi_1,\chi_2)$.
Next, the integration in an infinitesimal momentum layer
over the high spatial frequency part of the $\phi_1$-field is performed
in the background of the slowly varying fields $\phi_{0L},\phi_{1L},\chi_{1L},
\chi_{2L}$.  The accuracy of the result can be optimised by choosing
appropriate
$\phi_0,\phi_1$ dependence for the auxiliary fields $\chi_1$ and $\chi_2$.

The result of the integration is of non-polynomial nature.
Due to
the large mass of the $n=1$ mode the interaction is short ranged
and screens the infrared behaviour. In other words, the scale dependence
is given by the ultraviolet scaling relations for all length scales.
This allows to restrict the solution for up to the sixth order terms
of the local potential and to expand the renormalization group equation
in the interaction, (\ref{scaldiff}).
By this argument throughout this
paper we approximate the full potential energy density by a finite order
polynomial (namely, up to $\phi^4$ or $\phi^6$). The result of
an infinitesimal step
can be expressed then in terms of a set of coupled first order non-linear
differential equations. They describe the flow of the theory in
a restricted coupling space as the momentum scale bounding the support
of the heavy field ($\phi_1$) is lowered.

The effective light ($\phi_0$) theory is arrived at when the above bound
reaches zero, that is the $\phi_1$ integration being completed. It is
worthwhile to emphasize that $\phi_1$ does not fully vanish from the theory.
It is present as a constant background, the integration of the $\phi_0$-theory
(for instance the calculation of its effective potential) should be performed
in such background. This procedure clearly will react back on the final
$\phi_1$
dependence of this two-variable function. It might look just a curiosity, that
our two-step integration over the non-static modes allows us searching for a
non-trivial minimum in a two-variable $(\phi_0,\phi_1)$ domain. Usually it is
correct to assume that the absolute minimum lies for any temperature in the
$\phi_1=0$ slice. Still, it could occur in some segment of the coupling space,
that  a ground state which breaks invariance under $\tau$-translation is
realised.

\subsection{The generalised Hubbard-Stratonovich transformation
for the $\Phi^4$-theory}

In the following analysis the projection of the potential energy density
will be restricted to  dimension-6 operators.
The starting action at momentum scale $K$ is parametrized in the form:
\begin{eqnarray}
&
S[\phi_0 ,\phi_1]_K=\int d^3x [{1\over 2}(\partial_i\phi_0)^2+
|\partial_i\phi_1|^2+{1\over 2}A_1(K)\phi_0^2+{1\over 24}A_2(K)\phi_0^4
\nonumber\\
&
+((2\pi T)^2+B_1(K))|\phi_1|^2+{1\over 12}B_2(K)|\phi_1|^4
+C(K)\phi_0^2|\phi_1|^2\nonumber\\
&
+D_1(K)\phi_0^6+D_2(K)\phi_0^4|\phi_1|^2+D_3(K)\phi_0^2|\phi_1|^4+
D_4(K)|\phi_1|^6].
\label{kaction}
\end{eqnarray}
A two-step generalisation of the Hubbard-Stratonovich transformation
will be used, based on the following simple identities:
\begin{eqnarray}
&
\int_{C-i\infty}^{C+i\infty}d\chi_1\exp [{1\over 2}\chi_1^2-
\xi_1\chi_1|\phi_1|^2-\xi_2\chi_1|\phi_1|^4+{1\over 2}\xi_1^2|\phi_1|^4]
\nonumber\\
&
{}~~~~~~~~~~~~~~~~\sim\exp (-\xi_1\xi_2|\phi_1|^6
+{\cal O}(|\phi_1|^8),
\nonumber\\
&
\int_{C-i\infty}^{C+i\infty}d\chi_2\exp [{1\over 2}\chi_2^2-\xi_3
\chi_2|\phi_1|^2]\sim\exp (-{1\over 2}\xi_3^2|\phi_1|^4).
\label{HS}
\end{eqnarray}
Requiring the relations
\begin{equation}
\xi_1\xi_2=D_4,~~~~{1\over 2}\xi_3^2=D_3\phi_0^2+{1\over 12}B_2+\xi_2\chi_1
-{1\over 2}\xi_1^2
\label{ksidet}
\end{equation}
one can use the equalities (\ref{HS}) for rewriting the the action in extended
form (possible terms of ${\cal O}(|\phi_1|^8)$ are omitted):
\begin{eqnarray}
&
S[\phi_0 ,\phi_1]_K=\int d^3x \{{1\over 2}(\partial_i\phi_0)^2+
|\partial_i\phi_1|^2+{1\over 2}A_1(K)\phi_0^2+{1\over 24}A_2(K)\phi_0^4
\nonumber\\
&
+D_1(K)\phi_0^6-{1\over 2}(\chi_1^2+\chi_2^2)\nonumber\\
&
[(2\pi T)^2+B_1(K)+\xi_1\chi_1+\xi_3\chi_2+C(K)\phi_0^2+D(K)\phi_0^4]
|\phi_1|^2.\}
\label{extaction}
\end{eqnarray}

For the $\chi$-integrals we follow the usual approximation of saddle
point dominance. The long wavelength parts of both $\phi_0$ and $\phi_1$
are replaced by constants, which enables one to perform the integration over
the
short wavelength part of $\phi_1$ easily:
\begin{equation}
{1\over 2\pi^2}\int_k^Kdp~p^2\log (p^2+(2\pi T)^2+B_1+\tilde\chi_1+\tilde\chi_2
+C\phi_0^2+D_2\phi_0^4).
\label{oneloop}
\end{equation}

For the scaled quantities
\begin{equation}
\tilde\chi_1=\xi_1\chi_1,~~~\tilde\chi_2=\xi_3\chi_2
\end{equation}
one gets the following gap equations:
\begin{eqnarray}
&
\tilde\chi_1=(\xi_1^2+D_4\tilde\chi_2/ \xi_3^2)(|\phi_1|^2+I_1(\Delta )),
&
\tilde\chi_2=\xi_3^2(|\phi_1|^2+I_1(\Delta )).
\label{gapeq}
\end{eqnarray}
Here the notation $\Delta \equiv \tilde\chi_1+\tilde\chi_2+B_1+C\phi_0^2+
D_2\phi_0^4$ and the integrals
\begin{equation}
I_n(\Delta )={1\over 2\pi^2}\int_k^K{p^2dp\over (p^2+(2\pi T)^2+\Delta )^n}
\label{intn}
\end{equation}
have been introduced.

In the classical extended action (\ref{extaction}), evaluated with the
long wavelength parts of the fields the saddle point contribution can be
shown to differ from the classical expression
$D_3\phi_0^2|\phi_1|^4+B_1|\phi_1|^4/12+D_4|\phi_1|^6$
only in terms proportional to higher powers of $I_1$. They give negligible
contribution to the couplings at scale $k$, when the interval $(k,K)$
becomes infinitesimal.

Therefore, the running of the couplings comes exclusively from (\ref{oneloop}).
Its field-dependent part can be expanded into powers of
$|\phi_1|^2$ and $\phi_0^2$:
\begin{equation}
\Delta U=\sum_{n=1}^{\infty}{(-1)^{n+1}\over n}(\tilde\chi_1+\tilde\chi_2
+C\phi_0^2+D_2\phi_0^4)^nI_n(B_1).
\label{delu}
\end{equation}
Since each term is multiplied by an integral over the infinitesimal range
$(k,K)$, for the differential running that part of $\tilde\chi_1+\tilde\chi_2$
which is proportional to another similar integral $I_n$ does not play any role.
Combining the two equations of (\ref{gapeq}) and the definitions of $\xi_1^2,
\xi_3^2$, one can show that
\begin{equation}
\tilde\chi_1+\tilde\chi_2=2D_3\phi_0^2|\phi_1|^2+3D_4|\phi_4|^4+
{1\over 6}B_2|\phi_1|^4+{\cal O}(|\phi_1|^6)
\label{hiclass}
\end{equation}
is the relevant (classical) value to be used for the derivation of the
differential running equations of the various couplings. The
${\cal O}(|\phi_1|^6)$ contribution would produce a linearly divergent
term in the running of $D_4$. It should be omitted by the neglect
of the irrelevant $|\phi_1|^8$ term. The full set of evolution equations
appears in Appendix B. Here we discuss analytically the evolution of dimension
2 and 4 operators.

When one projects the potential onto a fourth order polynomial,
it is sufficient to keep only the first two terms of (\ref{delu}).
The most convenient is to introduce dimensionless variables by scaling
all three-dimensional couplings by appropriate powers of $2\pi T$:
\begin{eqnarray}
&
x={k\over 2\pi T},~~a_1={A_1\over (2\pi T)^2},~~b_1={B_1\over (2\pi T)^2},
\nonumber\\
&
{}~~a_2={A_2\over 2\pi T},~~b_2={B_2\over 2\pi T},~~c={C\over 2\pi T}.
\end{eqnarray}
Then the following simple set of coupled first order differential equations
arises:
\begin{eqnarray}
&
{da_1\over dx}=-{c\over \pi^2}{x^2\over x^2+1+b_1},~~
{db_1\over dx}=-{b_2\over 12\pi^2}{x^2\over x^2+1+b_1},\nonumber\\
&
{dc\over dx}={b_2c\over 12\pi^2}{x^2\over (x^2+1+b_1)^2},
{}~~{da_2\over dx}={6c^2\over \pi^2}{x^2\over (x^2+1+b_1)^2},\nonumber\\
&
{db_2\over dx}={b_2^2\over 12\pi^2}{x^2\over (x^2+1+b_1)^2}.
\label{scaldiff}
\end{eqnarray}

Comparing the right hand sides one easily finds some natural relationships
between the evolution of various couplings:
\begin{equation}
c=Q_1b_2,~~~a_2=72Q_1^2b_2+Q_2,~~~a_1=12Q_1b_1+Q_3,
\label{lincon}
\end{equation}
where $Q_i$ are integration constants. These relations imply that
the true task is the solution of the coupled evolution equations of $b_1$ and
$b_2$.

In the weak coupling regime, where also $b_1$ can be assumed in
perturbative sense to be small,
in the first approximation $b_1$ can be neglected in the denominator of
the right hand side of the equation for $b_2$ and can be represented by
an expansion of the denominator up to linear terms
on the right hand side of the equation of $b_1$ (One works
to quadratic order on the right hand sides of (\ref{scaldiff})).
Then, it is easy to find explicit analytic solutions:
\begin{eqnarray}
&
b_2(x)={24\pi^2(1+x^2)\over x+(1+x^2)(24\pi^2Q_4-\tan^{-1}x)},\nonumber\\
&
b_1(x)={1+x^2\over x+(1+x^2)(24\pi^2Q_4-\tan^{-1}x)}
[Q_5+2(\tan^{-1}x-x)].
\label{solphi}
\end{eqnarray}
The integration constants can be found from the intial conditions which are
the couplings at momentum scale $\Lambda$, arising from the 1-loop integration
over the $|n|\geq 2$ modes:
\begin{eqnarray}
&
a_1(x_{\Lambda})=b_1(x_{\Lambda})={m^2\over (2\pi T)^2}+{13\lambda
\over 96\pi^2}-{3\lambda\Lambda \over 16\pi^4 T},\nonumber\\
&
a_2(x_{\Lambda})=
{\lambda\over 2\pi},~~b_2(x_{\Lambda})={3\lambda\over 2\pi}(1+b\lambda),~~
c(x_{\Lambda})={\lambda\over 4\pi}(1+a\lambda).
\label{initcon}
\end{eqnarray}
The constants $Q_i$ are found with straightforward algebra when equating
(\ref{initcon})
to the large $x_{\Lambda}$ limiting values of the corresponding functions in
(\ref{lincon}) and (\ref{solphi}).

  The couplings reflecting the effect of $\phi_1$-fluctuations are given
by $a_i(0),b_i(0),$\\ $c(0)$. The couplings $a_i,c$ determine the effective
fluctuation dynamics of the static modes ($\phi_0$). $b_i$ characterize
the effective potential of constant $\phi_1$ configurations.
The assumption of the smallness of $B_2$ and $C$
which was essential for the consistent truncation of the
infinite set of RG-equations to (\ref{scaldiff})
is ensured by restricting $\lambda$ to small values. Then,
it is sufficient to give
the first few terms of the power series of different couplings with respect to
$\lambda$:
\begin{eqnarray}
&
b_2(0)\approx {3\lambda\over 2\pi}(1+\lambda(b-{1\over 32\pi^2})),
{}~~c(0)\approx {\lambda\over 4\pi}(1+\lambda (a-{1\over 32\pi^2})),\nonumber\\
&
a_2(0)\approx {\lambda\over 2\pi}(1-{3\lambda\over 32\pi^2}),\nonumber\\
&
b_1(0)\approx {m^2\over (2\pi T)^2}\bigl (1-{\lambda \over 32\pi^2}(1-
\lambda({1\over 32\pi^2}-b))\bigr )+{7\lambda\over 96\pi^2}
-{\lambda^2\over 16\pi^2}({7\over 192\pi^2}+{1\over 8\pi^4}+b)\nonumber\\
&
{}~~~~~~~~~~~~~-{\lambda\Lambda\over 8\pi^4T}(1-{\lambda\over 2}(b+{1\over
16\pi^2})),\nonumber\\
&
a_1(0)\approx {m^2\over (2\pi T)^2}\bigl (1-{\lambda\over 16\pi^2}(1+
\lambda(a-{1\over 32\pi^2}))\bigr )+{\lambda\over 96\pi^2}-
{\lambda^2\over 8\pi^2}({7\over 192\pi^2}+{1\over 8\pi^4}+a)
\nonumber\\
&
{}~~~~~~~~~~~~~~+{\lambda\Lambda\over 8\pi^4T}(-{1\over 2}+\lambda
(a+{1\over 16\pi^2})).
&
\label{pertnonpert}
\end{eqnarray}
\subsection{Justification of the truncation}

The solution of the approximate renormalization group equations
(\ref{scaldiff})
is based on the assumption $1+x^2>|b_1|$ which amounts to the inequalities
\begin{equation}
{m^2\over(2\pi T)^2}+{13\lambda\over96\pi^2}-{3\lambda\Lambda\over16\pi^4 T}
<1+{\Lambda^2\over(2\pi T)^2},
\label{irp}
\end{equation}
or
\begin{equation}
{m^2\over(2\pi T)^2}+{13\lambda\over96\pi^2}-{3\lambda\Lambda\over16\pi^4 T}
>-1-{\Lambda^2\over(2\pi T)^2}.
\label{irm}
\end{equation}

(\ref{irp}) can be taken for granted since its violation requires
\begin{equation}
0<T^2\biggl(4\pi^2-{13\lambda\over24}-{9\lambda^2\over56\pi^4}\biggr)<m^2,
\end{equation}
and $\Lambda=O(\lambda T)$.

(\ref{irm}) represents a more stringent condition,
\begin{equation}
m^2>-T^2\biggl(4\pi^2-{13\lambda\over24}-{9\lambda^2\over56\pi^4}\biggr).
\end{equation}
For $m^2=-O((2\pi T)^2)$ the $\phi_1$ system becomes critical for
$\phi_0=0$. But this infrared instability is misleading,
it comes from the parametrization of the effective theory only. The
coupling constants $a_1$, $a_2$ and $c$ were obtained by expanding the
effective
potential around $\phi_0=0$, so they are relevant in describing the
fluctuations in the symmetrical phase or in the vicinity of the
phase transition. For deeply broken symmetry systems the mass
term $C\phi_0^2$ stabilizes the $\phi_1$ system.
In this case the usefulness of the
integration of $\phi_1$ alone is questionable and the
$\phi_0-\phi_1$ systems should be treated together.

These arguments demonstrate that the numerical solution of the original
equation  (\ref{scaldiff}) would not provide in the weak coupling
regime quantitatively different information on the reduced action compared to
the approximation where the denominator is expanded in powers of $b_1$.

\subsection{Comparison with results of simplified treatments of the
$\phi_1$-integration}

The coefficients $a_i$ appearing in (\ref{pertnonpert})
 can be compared with less
sophisticated ways of integrating out $\phi_1$.

The simplest is the
1-loop integration with constant $\phi_0$ background, which is a version of the
hierarchical integration advocated previously within the 3D effective
theory of the electroweak phase transition \cite{jako1}.
The quadratic form in the Fourier-space is simply determined by
\begin{equation}
k^2+m^2+(2\pi T)^2+{13\lambda T^2\over 24}+
{\bar\lambda\phi_0^2\over 2}(1+a\lambda ),
\end{equation}
which leads to the following contribution to the effective $\phi_0$-theory:
\begin{equation}
{\bar\lambda\Lambda T\over 2\pi^2}(1+a\lambda ){1\over 2}\phi_0^2
-{T\over 6\pi}
(m^2+(2\pi T)^2+{13\lambda T^2\over 24}+{\bar\lambda\phi_0^2\over 2}
(1+a\lambda ))^{3/2}.
\label{nocsak}
\end{equation}
For $\bar\lambda\phi_0^2<<T^2$ one expands this expression in power
series with respect to $\lambda$ up to terms ${\cal O}(\lambda^2)$,
what is equivalent to retain terms only up to ${\cal O}(\phi_0^4)$.
In the expressions below also the leading ${\cal O}(m^2/T^2)$ corrections are
included. The results for the
coefficients $a_i$ are the following:
\begin{eqnarray}
&
a_1(0)={m^2\over (2\pi T)^2}(1-{\lambda\over 16\pi^2}(1+a\lambda)+
{13\lambda^2\over 3072\pi^4})+
{\lambda\over 96\pi^2}-{\lambda^2\over 8\pi^2}(a+{13\over 192\pi^2})\nonumber\\
&
{}~~~~~~~~~~~~~-{\lambda\Lambda\over 8\pi^4 T}({1\over 2}-a\lambda),\nonumber\\
&
a_2(0)={\lambda\over 2\pi}(1-{3\over 16\pi^2}\lambda
(1-{m^2\over 8\pi^2T^2})).
\label{loopm}
\end{eqnarray}
One instantly recognizes that the difference between (\ref{pertnonpert})
and (\ref{loopm}) of the resulting
couplings starts at ${\cal O}(\lambda^2)$.

One might improve upon the simplest 1-loop hierarchical integration
 by applying first
the HS-transformation, as described above. Essentially this amounts to
the evaluation of $I_n$ in (\ref{intn}) with $k=0,~K=\Lambda$, and keeping only
the $\phi_0$ background different from zero. The result is
\begin{eqnarray}
&
a_1(0)={m^2\over (2\pi T)^2}(1-{\lambda\over 16\pi^2}(1+a\lambda)
+{13\lambda^2\over 3072\pi^4})
+{\lambda\over 96\pi^2}-{\lambda^2\over 8\pi^2}(a+{1\over 192\pi^2})\nonumber\\
&
-{\lambda\Lambda\over 8\pi^4 T}({1\over 2}-\lambda (a-{1\over 32\pi^2})),
\label{saddlem}
\end{eqnarray}
while the expression of $a_2(0)$ agrees with the 1-loop (and also the RG)
result to ${\cal O}(\lambda^2)$.

The change in the shape of the effective $\Phi_0 - \Phi_1$ potential
depends on the renormalisation conditions. Since they were imposed
 at an intermediate stage of the integration, it is difficult to compare
the above result with, for instance, an ${\cal O}(\lambda^2)$ evaluation of the
effective potential without reduction. At present the best one can do is to
evaluate the phase transition temperature {\it a la Landau} from the vanishing
of the finite part of $a_1(0)$ in the different approximations.
The original estimate of Linde and Kirzhnitz \cite{kirznitz} receives an
${\cal O}(\lambda^0)$ correction as follows:
\begin{equation}
-{\lambda\over 24m^2}T_c^2= 1+0.0596\lambda~(RG),~1+0.1187\lambda
{}~(pert.),~ 1+0.0427\lambda(saddle),
\label{compare}
\end{equation}
as one calculates it from eqs.
(\ref{pertnonpert},\ref{loopm},\ref{saddlem}), respectively.
This can be compared
with the critical temperature calculated in the same way
from the effective static $\Phi_0$
theory arrived at with the reduction peformed
 on 2-loop level \cite{jako3}: $1+0.0462\lambda$.
This shows that the saddle point integration in the present context
is "equivalent" to the 2-loop calculation, while the non-perturbative
integration "sums up" the two- and higher loop contributions into a reasonably
nearby estimate.

\subsection{Non-perturbative $A_0$-integration in the effective electroweak
theory}

The integration of the non-static modes has provided a new distinct
screening scale, the Debye-mass of $A_0$, (\ref{ewdefmass}). This mass has
been shown in perturbative treatments to be consistently larger, near the
phase transition point, than the effective masses of the magnetic vector-
and Higgs-field fluctuations. Based on this observation a treatment
fully analogous to that of $\Phi_1$ can be envisaged also for $A_0$.

For the simplicity of  presentation we restrict the potential energy density
to its quartic projection. In Appendix B we show, that the results
are stable, when the projection is extended to dimension 6 operators.

Then the Lagrangian of the theory at scale $K$ can be
parametrized as
\begin{eqnarray}
&
L={1\over 4}F_{ij}F_{ij}+{1\over 2}(D_i\phi)^{\dagger}(D_i\phi)+{1\over 2}
(D_iA_0)^2+{1\over 2}A_1(K)\phi^{\dagger}\phi+
{1\over 24}A_2(K)(\phi^{\dagger}\phi)^2
\nonumber\\
&
{}~~~~~~~~~~~~~~~~+{1\over 2}B_1(K)A_0^2+{1\over 2}B_2(K)(A_0^2)^2+
C(K)\phi^{\dagger}\phi A_0^2.
\end{eqnarray}
After the HS-transformation is applied to the quartic $A_0$-potential, the
Lagrangian density gets the form:
\begin{eqnarray}
&
L={1\over 4}F_{ij}F_{ij}+{1\over 2}(D_i\phi)^{\dagger}(D_i\phi)+{1\over 2}
(D_iA_0)^2+{1\over 2}A_1\phi^{\dagger}\phi+{1\over 24}
A_2(\phi^{\dagger}\phi)^2\nonumber\\
&
{}~~~~~~~~-{1\over 8B_2}\chi^2+{1\over 2}(B_1+\chi)A_0^2+
C\phi^{\dagger}\phi A_0^2.
\end{eqnarray}
The integration over the Fourier components of $A_0$ belonging to
 the $(k,K)$-layer
contributes to the potential energy density
\begin{equation}
{3\over 4\pi^2}\int_k^K dpp^2
\log (p^2+B_1+\chi+2C\phi^{\dagger}\phi ).
\end{equation}
The omission of the magnetic vector potential from the background, in which
this expression has been evaluated corresponds to neglecting wave
function renormalisation effects.

The optimisation of the HS-transformation is achieved through  the choice of
$\chi$ according to the gap-equation:
\begin{equation}
-{\chi\over 4B_2}+{1\over 2}A_0^2+{3\over 4\pi^2}\int_k^K dpp^2
{1\over p^2+B_1+\chi+2C\phi^{\dagger}\phi }=0.
\end{equation}
It has the following iterative solution to quadratic order in the couplings:
\begin{equation}
\chi_0=2B_2(A_0^2+3I_1),~~~\chi_1=-4B_2I_2(2C\phi^{\dagger}\phi+
2B_2(A_0^2+3I_1)),
\end{equation}
where the notation
\begin{equation}
I_n={1\over 2\pi^2}\int_k^Kdpp^2{1\over (p^2+B_1)^n}
\end{equation}
has been used.

In this case it seems to be more convenient  to scale dimensionfull quantities
by appropriate powers of $m_D=\sqrt{5/6}gT$.
On this scale we expect $b_1\equiv B_1/m_D^2=1+{\cal O}(g)$.
With this assumption
we break up $b_1$ into $1+\bar b_1$. $\bar b_1$ is expected to sum up higher
order contributions, but not to exceed the order of magnitude
of $g$.
Repeating the same steps as in case of the
$\phi^4$ theory one arrives at the following set of renormalisation group
equations:
\begin{eqnarray}
&
{da_1\over dx}=-{3c\over \pi^2}{x^2\over x^2+1+\bar b_1},~~
{d\bar b_1\over dx}=-{3b_2\over \pi^2}{x^2\over x^2+1+\bar b_1},\nonumber\\
&
{da_2\over dx}={36c^2\over \pi^2}{x^2\over (x^2+1+\bar b_1)^2},
{}~~{db_2\over dx}={3b_2^2\over \pi^2}{x^2\over (x^2+1+\bar b_1)^2},\nonumber\\
&
{dc\over dx}={3b_2c\over \pi^2}{x^2\over (x^2+1+\bar b_1)^2}.
\label{eqa0}
\end{eqnarray}
Like in the previous model, again some natural relations can be found
between the quadratic couplings, as well as the quartic ones:
\begin{equation}
c=Q_1b_2,~~~a_1=Q_1b_1+Q_2,~~~a_2=12Q_1^2b_2+Q_3,
\end{equation}
therefore the problem is again reduced to the solution of the coupled
set of the two equations for $\bar b_1$ and $b_2$.
With the replacement
\begin{equation}
b_2={\bar b_2\over 36}
\end{equation}
in (\ref{eqa0}) one finds a form of the differential equations for $\bar b_1,
\bar b_2$ which is identical to (\ref{scaldiff}). It has the same solution
with some constants of integrations, which can be determined from the initial
conditions set at $x_{\Lambda}=\Lambda /m_D$:
\begin{eqnarray}
&
a_1(x_{\Lambda})={6\over 5}({\hat m^2\over (gT)^2}+({3\over 16}+
{\lambda\over 12g^2}) -\sqrt{{5\over 6}}({9\over 4}+{\lambda\over g^2})
{gx_{\Lambda}\over 2\pi^2 }),
{}~~\bar b_1(x_{\Lambda})=-{\sqrt{30}gx_{\Lambda}\over 2\pi^2 },\nonumber\\
&
a_2(x_{\Lambda})=\sqrt{{6\over 5}}{\hat\lambda\over g},
{}~~c(x_{\Lambda})=\sqrt{{6\over 5}}{g\over 8},
{}~~b_2(x_{\Lambda})=\sqrt{{6\over 5}}{17g^3\over 96\pi^2}.
\end{eqnarray}

The analytic solutions are given with help of (\ref{solphi}), if on the
basis of the assumption $\bar b_1<<1$ the $A_0$-propagators in (\ref{eqa0})
are expanded in powers of $\bar b_1$, keeping on the right hand sides
 terms up to quadratic order in the couplings.

After some ugly  algebra one arrives at the following expressions for the
effective couplings arising from the RG-improved integration over $A_0$:
\begin{eqnarray}
&
b_2(0)-b_2(x_{\Lambda})=-{289g^6\over 1024\pi^5}{1\over 1+\sqrt{{6\over 5}}
{17g^3\over 128\pi^3}}\nonumber\\
&
c(0)-c(x_{\Lambda})=-{51g^4\over 2560\pi^3}{1\over 1+\sqrt{{6\over 5}}
{17g^3\over 128\pi^3}}\nonumber\\
&
a_2(0)-a_2(x_{\Lambda})=-{27g^2\over 160\pi}{1\over 1+\sqrt{{6\over 5}}
{17g^3\over 128\pi^3}}\nonumber\\
&
\bar b_1(0)-\bar b_1(x_{\Lambda})=\sqrt{{6\over 5}}{17g^3\over 32\pi^4}
[(1+\sqrt{{6\over 5}}{5g\over 8\pi})x_{\Lambda}-{\pi\over 2}(1+\sqrt{{6\over
5}}
{5g\over \pi^3})]{1\over 1+\sqrt{{6\over 5}}
{17g^3\over 128\pi^3}}\nonumber\\
&
a_1(0)-a_1(x_{\Lambda})=\sqrt{{6\over 5}}{3g\over 8\pi^2}
[(1+\sqrt{{6\over 5}}{5g\over 8\pi})x_{\Lambda}-{\pi\over 2}(1+\sqrt{{6\over
5}}
{5g\over \pi^3})]{1\over 1+\sqrt{{6\over 5}}
{17g^3\over 128\pi^3}}
\label{3dhiggscoupl}
\end{eqnarray}
In all couplings in the physical $g$-region ($\sim 2/3$)
the common denominator can be omitted since it gives $\sim 1/1000$ relative
correction.

The expression of $\bar b_1$ receives  ${\cal O}(g^3)$ corrections,
justifying the starting assumption of $\bar b_1<<1$. In the common square
bracket of the expressions of $a_1(0)-a_1(x_{\Lambda}$) and $\bar
b_1(0)-\bar b_1(x_{\Lambda}$) the ${\cal O}(g)$ corrections
are of cca. 15\% relative importance in the interesting $g$-range.

For comparison we quote  the corrections to the potential energy density
from the one-step 1-loop integration of $A_0$ \cite {jako1}:
\begin{eqnarray}
&
(a_1(0)-a_1(x_{\Lambda}))[{\rm 1-loop}]=-\sqrt{{6\over 5}}{3g\over 16\pi}+
\sqrt{{6\over 5}}{3g\over
8\pi^2}x_{\Lambda},\nonumber\\
&
(a_2(0)-a_2(x_{\Lambda}))[{\rm 1-loop}]=-{27g^2\over 160\pi}.
\end{eqnarray}
One recognizes that the leading  corrections to the
scaled mass parameter, as calculated with both methods, coincide.
The non-trivial result is the next term,
which appears only in (\ref{3dhiggscoupl})
and mimicks the effect of a reduction on the 2-loop level.
We conclude that the improved
integration leads to higher order differences in the couplings. They might
influence  the result of the numerical simulation of the resulting
gauged Higgs system appreciably, since their contribution represents a
10-15\%  correction relative to the 1-loop $A_0$-integration.

\section{ Discussion and conclusions}

The derivation of an effective theory for three dimensional static
fields is presented in this paper by means of different approximations.
The idea of using effective models originates from combining
different methods in dealing with different
degrees of freedom.

At high temperature the natural strategy is
to eliminate first the non-static modes with the simplest one-loop
approximation. The more delicate problem of infrared
sensitive static modes is dealt with in the effective model where the
dynamics of the non-static modes is incorporated into the effective
vertices. Even the simplest leading order perturbative
solution of the effective theory represents a resummation of the
thermal mass.

This well-known scheme is a rudimentary application of the scheme of the
renormalization group where the contributions of the modes which have
been eliminated are taken into account in the next elimination step.
In order to improve the effective model we pursued more systematically the
renormalisation group
strategy in the elimination of some massive non-static modes in
the 1-component scalar model
and of the $A_0$ components of the vector fields in the SU(2) Higgs model.

In the scalar model the $n>1$ Matsubara modes were integrated out in the
first step in the
one-loop, independent mode approximation resulting in an effective
theory for the $\phi_1$ and $\phi_0$ modes. By this we ensure that
in the second step when the
$n=1$ modes are eliminated the interactions between the $n>1$ and
the $n=1$ modes are retained. Since $\phi_1$ is the lightest heavy
mode we performed a partial resummation of the perturbation expansion
during its elimination. This consisted of applying a saddle point
approximation for a Hubbard-Stratonovich  auxiliary field
representing nonlinearities of the ineratctions of the $\phi_1$ field,
and solving
the renormalization group equations for the relevant operators.

Higher loop contributions in the solution (\ref{solphi})
are generated by irrelevant, higher order operators when projected back
onto the relevant ones during the one-loop elimination step.
For the solution of the renormalization
group equation such  UV divergence structure is expected which is different
from that of a systematic multi-loop elimination.
The difference between the one-loop and the renormalization group solution
manifests itself in the higher loop contribution,
${\lambda^2\Lambda\over2(2\pi)^6}$, in (\ref{pertnonpert}).

In the framework of the renormalized perturbation expansion all UV
divergent contributions are treated as small perturbations because they are
multiplied by a positive power of the coupling constant. Our renormalization
group refers to the bare theory and does not distinguish between the
renormalized part of the lagrangian and the counterterms. Thus it is not
obvious to what extent the results obtained with the help of this equation
would correspond to a partial resummation of renormalized graphs.

It turns out that the correspondence between the partial resummation of the
renormalized perturbation expansion and the solution of the renormalization
group equation can actually be maintained. To see this we start
with the observation that the scheme of the renormalized
perturbation expansion could be violated if the inequality
\begin{equation}
{3\lambda\Lambda T\over4\pi^4}>| m^2+13\lambda T^2/24|
\end{equation}
is fulfilled, i.e. when the mass counterterm is larger than the finite part.
For $m^2<0$ and $\lambda\approx0$ this gives
\begin{equation}
x={\Lambda\over2\pi T}>-{2\pi m^2\over3\lambda T^2}.
\end{equation}
Since $x^2>x>1$ one can treat the counterterm as small beside
 the {\it kinetic} energy and the expansion in the counterterms and the finite
parts is justified by the same stroke. (This result holds in four dimensions,
too, since the mass counterterm, though being quadratically divergent, is
suppressed by $\lambda$ when compared with $\Lambda^2$.)

According to the argument presented above
the UV divergent parts can be treated perturbatively
for $\lambda<<1$. Since $\lambda$ has only finite renormalization this
inequality remains valid in the UV regime. Still, the "small" UV divergent part
of the solution of the renormalization group equation makes the
renormalization of the theory more involved. If the effective theory
is derived and solved in the n-loop approximation then the general proof
of the renormalizabiltiy allows to eliminate the cut-off. When the
derivation and the solution of the effective theory are based on different
approximations then the divergent parts of the bare coupling constants
of the effective theory do not cancel the divergences in the solution.
In this manner unless the derivation and the solution of the effective theory
match the resulting model is, strictly speaking, non-renormalizable.

This can be illustrated by the inspection of (\ref{pertnonpert})
and (\ref{nocsak}). The one-loop integration of $\phi_1$ generates
ultraviolet divergences in (\ref{nocsak}) which are exactly cancelled by
subsequent
one-loop integration of $\phi_0$. But when, inconsistently, tree-level
approximation is used for $\phi_0$ then the effective potential remains
cut-off dependent. In an analogous fashion the "bare" mass, $a_1(0)$
for $\phi_0$ in (\ref{pertnonpert}) was obtained by the renormalisation group
improved perturbation expansion and its cut-off dependence will not be
compensated for by its straight or any improved one-loop integration.
In the same time the mass of $\phi_1$, $b_1(0)$, displays different ultraviolet
divergences which remain unchanged after the integration over $\phi_0$.
At the end one finds different cut-off dependence for the masses
of $\phi_0$, $\phi_1$ and these infrared quantities can not be kept
simultaneously finite.

A pragmatic way out from this problem is offered by the careful
examination of the cut-off dependences. If the derivation and the solution
differ in higher order of the perturbation expansion then the
mismatch is $O(\lambda^n\Lambda T)$, $n>1$. The cut-off should be large
enough compared to the light mass scale,
$m_\ell=\sqrt{m^2+{13\lambda T^2\over24}}$, and small enough to keep this
mismatch smaller than the characteristic heavy mass square, $m_h=2\pi T$.
If the expected scaling behaviour can be established for
\begin{equation}
m_\ell<\Lambda<{m_h^2\over\lambda^nT},
\label{wind}
\end{equation}
then the given solution of the effective theory is acceptable in the
scaling window (\ref{wind}).

It might happen that in the relevant cut-off range only a somewhat
more modest approach can be followed, corresponding to a phenomenological
scaling law for the mass squares,
\begin{equation}
m^2(k=0,T,\Lambda)=m^2(T)+T\Lambda f(T/\Lambda).
\end{equation}
If such relation is found to be fulfillled over an extended cut-off range
 then one might conjecture that it belongs to a certain
renormalizable solution of the full theory.

This incertitude in the use of the divergent parts of the effective theory
is also a major problem in extracting physical results for the finite
temperature phase transition of the SU(2) Higgs model when its 3 dimensional
effective representation is being used, for instance, in the calculation
of $T_c$. The unscaled form determined in this paper looks like
\begin{eqnarray}
&
S_{eff}=\int d^3x[{1\over 4}F_{ij}^aF_{ij}^a+{1\over 2}
(D_i\phi)^{\dagger}(D_i\phi){1\over 2}+m_3^2(T)\phi^{\dagger}\phi+
{1\over 24}\lambda_3(T)(\phi^{\dagger}\phi)^2],\nonumber\\
&
m_3^2(T)=\hat m^2+[{3\over 16}g^2+{1\over 12}\lambda-{3g^2\over 16\pi}
{\sqrt{5\over 6}}(1+{\sqrt{6\over 5}}{5g\over \pi^3}){1\over
1+{\sqrt{6\over 5}}{17g^3\over 128\pi^3}}]T^2\nonumber\\
&
{}~~~~~-{\Lambda T\over 2\pi^2}[{9\over 4}g^2+\lambda-{3\over 4}g^2
(1+{\sqrt{6\over 5}}{5g\over 8\pi}){1\over
1+{\sqrt{6\over 5}}{17g^3\over 128\pi^3}}],\nonumber\\
&
\lambda_3(T)=[\hat\lambda-{\sqrt{5\over 6}}{27g^2\over 160\pi}){1\over
1+{\sqrt{6\over 5}}{17g^3\over 128\pi^3}}]T.
\end{eqnarray}

It is very important for non-perturbative integrations of this effective model
to test, how well the expected cut-off dependence (displayed in the second line
of the expression of $m_3^2$) fits the actually observed scaling behaviour.
The theoretical inaccuracy of the result is assessed by discussing the
sensitivity of the critical data ($T_c$, etc.) to the ${\cal O}(g^4T^2)$ and
${\cal O}(g^3\Lambda T)$ terms in $m_3^2(T)$.
\vskip .5truecm

\subsection*{Appendix A}

\def\theequation{A\arabic{equation}}
\setcounter{equation}{0}

{\it 1-loop integration of the $|n|\geq 2$ Matsubara modes}
\vskip .3truecm
The  1-loop integration over the $n\neq 0,\pm 1$ Matsubara modes with
space-independent, $\tau$-dependent background
\begin{eqnarray}
&
\Phi (x,\tau )=\Phi (\tau )+\phi (x,\tau )\nonumber\\
&
\Phi (\tau )=\Phi_0+\Phi_1\exp (i\omega \tau)
+\Phi_{1}^*\exp (-i\omega\tau ),~~~~\omega =2\pi T,\nonumber\\
&
\phi (x,\tau )=\sum_{n\neq 0,\pm 1}\phi_n(x)\exp (i\omega n\tau ).
\end{eqnarray}
 starts by writing explicitely the action up to terms
quadratic in the $|n|\geq 2$ Matsubara modes:
\begin{eqnarray}
&
  S=\beta V \bigl [{1\over 2} m^2(\Phi_0^2+2|\Phi_1|^2)+{1\over 24}\lambda
(\Phi_0^4+12\Phi_0^2|\Phi_1|^2+6|\Phi_1|^4)\bigr ]\nonumber\\
&
+\beta\bigl [ {1\over 2}m^2+{\lambda\over 4}(\Phi_0^2+2|\Phi_1|^2)\bigr ]
\sum_{k,n}^{'}
\phi_{k,n}^*\phi_{k,n}\nonumber\\
&
+\beta{\lambda\over 2}\Phi_0\sum_{k,n}^{'}(\Phi_{1}^*
\phi_{k,n}^*\phi_{k,n+1}+\Phi_1\phi_{k,n}\phi_{k,n+1}^*)\nonumber\\
&
{}~~~~~~+\beta{\lambda\over 4}\sum_{k,n}^{'}(\Phi_1^2\phi_{k,n}\phi_{k,n+2}^*
+\Phi_{1}^{*2}\phi_{k,n}^*\phi_{k,n+2})\nonumber\\
&
{}~~~~~~+{\beta\lambda\sqrt{V}\over 6}[3\Phi_0\Phi_1^2\phi_{0,-2}+3\Phi_0
\Phi_1^{*2}\phi_{0,2}+\Phi_1^3\phi_{0,-3}+\Phi_1^{*3}\phi_{0,3}]+
{\cal O}(\phi^3).
\end{eqnarray}
{}From the primed sums
the modes $n=0,\pm 1$ are always omitted.

The matrix of the quadratic form splits up into two identical disjoint blocs
for fixed $k$, one for Matsubara modes $n\geq 2$ and the other for $n\leq -2$:
\begin{equation}
M(k)=\left(\matrix{
k^2+\omega_2^2+M^2
&\lambda\Phi_1^*\Phi_0&{\lambda\over 2}\Phi_1^{*2}&0&0&...\cr
\lambda\Phi_1\Phi_0&k^2+\omega_3^2+M^2
&\lambda\Phi_1^{*}\Phi_0&...&0&...\cr
{\lambda\over 2}\Phi_1^2&\lambda\Phi_1\Phi_0&k^2+\omega_4^2+M^2
&...&...&0\cr
0&...&...&...&...&...\cr}\right),
\label{fluctmatr}
\end{equation}
where
$M^2=m^2+{\lambda\over 2}(\Phi_0^2+2|\Phi_1|^2)$.

The contribution to the potential part of the effective action is of the form
\begin{equation}
\sum_k\log\det M(k)-\beta N_n^* M(0)_{nm}^{-1}N_m,
\label{fluctcon}
\end{equation}
with
\begin{equation}
N_n=\lambda\sqrt{V}\left (\matrix {
{1\over 2}\Phi_0\Phi_1^{*2}\cr
{1\over 6}\Phi_1^{*3}\cr
0\cr
.\cr
.\cr}\right).
\end{equation}
The $0$ argument of $M$ in the second term of (\ref{fluctcon})
tells that the inverse of the
matrix (\ref{fluctmatr}) should be evaluated at zero
spatial momentum. Since we are
interested in the reduced action up to terms of ${\cal O}(\Phi^6)$,
it is sufficient to keep only the diagonal terms in this inverse matrix
\begin{equation}
M(0)^{-1}\sim \left(\matrix{
(m^2+\omega_2^2)^{-1}&0&..&..\cr
0&(m^2+\omega_3^2)^{-1}&..&..\cr
..&..&..&..\cr}\right).
\end{equation}
The contribution of this term
to the sixth power part of the potential is calculated readily:
\begin{equation}
\Delta U_6^{(1)}=-\beta V\lambda^2\bigl ({1\over 4}\Phi_0^2|\Phi_1|^4{1\over
m^2+\omega_2^2}+{1\over 36}|\Phi_1|^6{1\over m^2+\omega_3^2}\bigr ).
\label{delu6}
\end{equation}

The expression of the first term in (\ref{fluctcon}), expanded
up to sixth power terms in
$\Phi_0$ and $\Phi_1$ reads
\begin{eqnarray}
&
\sum_k\log\det M(k)\simeq\sum_{k,n}^{'}\bigl\{ {M^2\over k^2+\omega_n^2}-
{M^4\over 2(k^2+\omega_n^2)^2}+{M^6\over 3(k^2+\omega_n^2)^3}\nonumber\\
&
-\lambda^2\Phi_0^2|\Phi_1|^2{1\over (k^2+\omega_n^2)(k^2+\omega_{n+1}^2)}
-{\lambda^2|\Phi_1|^4\over 4(k^2+\omega_n^2)
(k^2+\omega_{n+2}^2)}
-{\lambda^3\Phi_0^2|\Phi_1|^4\over (k^2+\omega_n^2)(k^2+\omega_{n+1}^2)
(k^2+\omega_{n+2}^2)}
\nonumber\\
&
+\lambda^2\Phi_0^2|\Phi_1|^2M^2({1\over (k^2+\omega_n^2)
(k^2+\omega_{n+1}^2)^2}+{1\over (k^2+\omega_n^2)^2(k^2+\omega_{n+1}^2)})
\nonumber\\
&
+{\lambda^2\over 4}|\Phi_1|^4M^2({1\over (k^2+\omega_n^2)^2
(k^2+\omega_{n+2}^2)}+{1\over (k^2+\omega_n^2)(k^2+\omega_{n+2}^2)^2})\bigr\}.
\label{ksum}
\end{eqnarray}

Direct evaluation of the contributions proportional to $\Phi^2$ leads to
\begin{equation}
\Delta U_2=\beta V{1\over 2}\lambda(\Phi_0^2+2|\Phi_1|^2)
({\Lambda^2\over 16\pi^2}
+{13T^2\over 24}-{3\Lambda T\over 4\pi^2}).
\end{equation}

For the evaluation of the fourth and sixth power contributions from
(\ref{ksum})
it is very
convenient to make use of the "mixed" or "Saclay"-representation of the n-sums
\cite{pisarski}:
\begin{eqnarray}
&
\Delta U_4={\beta V\over 2}\int {d^3k\over (2\pi)^3}\int_0^{\beta}d\tau
(-{M^4\over 2}-\lambda\Phi_0^2|\Phi_1|^2e^{i\omega\tau}-
{\lambda^2\over 4}|\Phi_1|^4e^{2i\omega\tau})G'(\tau )^2,\nonumber\\
&
\Delta U_6^{(2)}=\beta V\int {d^3k\over (2\pi)^3}\int_0^{\beta}d\tau_1
G'(\tau_1)\int_0^{\beta}d\tau_2G^{'}(\tau_1+\tau_2 )G^{'}(\tau_2 )\nonumber\\
&
\times
[{M^6\over 3}+\lambda^2\Phi_0^2|\Phi_1|^2M^2
e^{i\omega\tau_1}+{\lambda^2\over 4}|\Phi_1|^4M^2e^{2i\omega\tau_1}
\nonumber\\
&
{}~~~~~~~~~~~~~-\lambda^3\Phi_0^2|\Phi_1|^4e^{i\omega(\tau_1+2\tau_2)}].
\end{eqnarray}
Here the mixed finite temperature propagator is of the form
\begin{eqnarray}
&
G^{'}(\tau )=T\sum_{j\neq 0,\pm 1}e^{-ij\omega\tau}
{1\over k^2+\omega_j^2}\equiv  G(\tau ,k)-{T\over k^2}-{T\over k^2+\omega^2}
(e^{i\omega\tau}+e^{-i\omega\tau})\nonumber\\
&
G(\tau ,k)={1\over 2k}\sum_{s=\pm}f_s(k)e^{-sk\tau },~~~~0\leq\tau\leq\beta,
\nonumber\\
&
f_-=n_k,~~~f_+=1+n_k,~~~n_k=(e^{\beta k}-1)^{-1}.
\end{eqnarray}

We elaborate on the computation of the quartic part, where the following
relation, valid for any $l\neq 0$ can be exploited:
\begin{eqnarray}
&
\int_0^{\beta}d\tau e^{il\omega\tau}G'(\tau )^2={1+2n_k\over
k(4k^2+l^2\omega^2) }-
{2T\over k^2+\omega^2}({1\over k^2+(l-1)^2\omega^2}+{1\over
k^2+(l+1)^2\omega^2}) \nonumber\\
&
{}~~~~-{2T\over k^2(k^2+l^2\omega^2)}
+{2T\over k^2(k^2+\omega^2)}(\delta_{l,1}+\delta_{l,-1})+
{T\over (k^2+\omega^2)^2}(\delta_{l,2}+\delta_{l,-2}).
\label{int1}
\end{eqnarray}
Upon the application of (\ref{int1}) in the expression of $\Delta U_4$
one arrives at
the following representation containing only 1-variable integrals:
\begin{eqnarray}
&
\Delta U_4=
-{\lambda^2\over 32\pi^2}(\Phi_0^2+2|\Phi_1|^2)^2\int dx x^2
\bigl [{1\over 4x^3}(1+2{1\over e^x-1})+{e^x\over 2x^2(e^x-1)^2}-{1\over
x^4}
\nonumber\\
&
{}~~~~~~~~~-{2\over (x^2+4\pi^2)^2}\bigr ]\nonumber\\
&
-\beta V{\lambda^2\over 4\pi^2}\Phi_0^2|\Phi_1|^2\int dx x^2\bigl [{
1\over 4x(x^2+\pi^2)}(1+{2\over e^x-1})-{2\over x^2(x^2+4\pi^2)}\nonumber\\
&
{}~~~~~~~~~-{2\over
(x^2+4\pi^2) (x^2+16\pi^2)}\bigr ]\nonumber\\
&
-{\lambda^2\over 16\pi^2}|\Phi_1|^4\int dx x^2\bigl [{1\over
4x(x^2+4\pi^2)}(1+2{1\over e^x-1})-{2\over x^2(x^2+16\pi^2)}-{1\over
(x^2+4\pi^2)^2}\nonumber\\
&
{}~~~~~~~~ -{2\over (x^2+4\pi^2)(x^2+36\pi^2)}\bigr ].
\label{delu4}
\end{eqnarray}
Here the notation $x=\beta k$ has been introduced, and for the
divergent integrals a sharp cut-off $\Lambda\beta$ is understood.

{}From the large $x$ behavior of the integrand one easily extracts the
divergent piece of $\Delta U_4$:
\begin{eqnarray}
&
\Delta U_{4,div}=-\beta V{\lambda^2\over 128\pi^2}\ln{\Lambda\over \mu}
[\Phi_0^4+12\Phi_0^2|\Phi_1|^2+6|\Phi_1|^4]\nonumber\\
&
={V\over 24}\int_0^{\beta} d\tau \Phi (\tau)^4(-{3\lambda^2\over
16\pi^2}\ln{\Lambda\over \mu}).
\label{deludiv}
\end{eqnarray}
This result reproduces the usual 1-loop counterterm of the
$\phi^4$-theory correctly.

In (\ref{delu4}) there are 3 integrals which can be
evaluated only numerically:
\begin{eqnarray}
&
{1\over 2\pi^2}\int_0^{\Lambda\beta}dx[{1\over 4x}{e^x+1\over e^x-1}+
{e^x\over 2(e^x-1)^2}-{1\over x^2}]={1\over
8\pi^2}\ln{\Lambda\beta\over \pi}-1.4133\times 10^{-2},\nonumber\\
&
{1\over 8\pi^2}\int_0^{\infty}dx{x\over x^2+\pi^2}{1\over e^x-1}=
1.7121\times 10^{-3},\nonumber\\
&
{1\over 8\pi^2}\int_0^{\infty}dx {x\over x^2+4\pi^2}{1\over e^x-1}=
4.8875\times 10^{-4}.
\label{numint}
\end{eqnarray}

The logarithmic divergence of the first integral of (\ref{numint}) contributes
to (\ref{deludiv}), and is absorbed eventually by the counterterm
defined through the renormalisation condition:
\begin{equation}
{\partial^4U_4\over\partial\Phi_0^4}|_{\Phi_1=0}=\lambda_R.
\end{equation}
This leads to the appearence of finite corrections to the couplings of
the other quartic terms of the potential, which are of ${\cal O}(\lambda_R)$
relative to their tree-level values:
\begin{eqnarray}
&
\Delta U_4={\lambda_R\beta\over 24}\int
d^3x[\Phi_0^4(x)+12(1+a\lambda_R)\Phi_0^2|\Phi_1|^2+
6(1+b\lambda_R)|\Phi_1|^4 ],\nonumber\\
&
a=3.56\times 10^{-3},~~~~~b=3.17\times 10^{-3}.
\end{eqnarray}

The evaluation of $\Delta U^{(2)}_6$ is a quite tedious procedure. The easiest
way to work out the complicated algebraic expressions
of its integrand by making
use of a symbolic programming package (MATHEMATICA). It is worthwhile to
emphasize the importance of the correct implementation of the periodicity of
$G'(\tau_1+\tau_2)$ on the square-interval
$0\leq\tau_1\leq\beta,~0\leq\tau_2\leq\beta.$ The result of the
$\tau$-integrations could be expressed in analytic form, though its
cumbersome form does not give any valuable insight. The radial $k$-integration
has to be performed numerically (similarly to the case of $\Delta U_4$, given
in (\ref{delu4})). The integrand is finite both in
the infrared and the ultraviolet
domain, though separate terms might display singularities.

The final expression with explicit numerical coefficients looks like
\begin{eqnarray}
&
\Delta U_6={\beta V\over T^2}[5.40183~10^{-6}
M^6+2.7566~10^{-5}\lambda^2\Phi^2|\Phi_1|^2M^2\nonumber\\
&
+6.6586~10^{-6}
\lambda^2|\Phi_1|^4M^2-1.6991~10^{-5}\lambda^3\Phi_0^2|\Phi_1|^4].
\end{eqnarray}
This form also provides ${\cal O}({m^2\over T^2})$ corrections to the
coefficients of lower dimensional operators. All coefficients of dimension-6
operators are ${\cal O}(\lambda^3)$. In addition the numerical coefficients
are 2 orders of magnitude smaller than in front of the ${\cal O}(\lambda^2)$
terms in $\Delta U_4$. For this reason the starting values of these
terms one can rightfully represented by  (\ref{delu6}).

\subsection*{Appendix B}

\def\theequation{B\arabic{equation}}
\setcounter{equation}{0}

\vskip .5truecm
{\it 1. Flow-equations for the couplings of all dimension 6 non-derivative
operators of the
$\Phi^4$-theory upon integration over $\Phi_1$}
\vskip .3truecm
The couplings appearing in the parametrisation of the action at scale $K$
(\ref{kaction}) will flow following the equations, which are derived
identically to (\ref{scaldiff}), just including all operators up to dimension
6:
\begin{eqnarray}
&
{da_1\over dx}=-{c\over \pi^2}{x^2\over x^2+1+b_1},~~
{db_1\over dx}=-{b_2\over 12\pi^2}{x^2\over x^2+1+b_1},\nonumber\\
&
{dc\over dx}={b_2c\over 12\pi^2}{x^2\over (x^2+1+b_1)^2}-{D_3\over \pi^2}
{x^2\over x^2+1+b_1},\nonumber\\
&
{da_2\over dx}={6c^2\over \pi^2}{x^2\over (x^2+1+b_1)^2}-{12D_2\over \pi^2}
{x^2\over x^2+1+b_1},\nonumber\\
&
{db_2\over dx}={b_2^2\over 12\pi^2}{x^2\over (x^2+1+b_1)^2}-{18D_4\over \pi^2}
{x^2\over x^2+1+b_1},\nonumber\\
&
{dD_1\over dx}={D_2c\over 2\pi^2}{x^2\over (x^2+1+b_1)^2}-{c^3\over 6\pi^2}
{x^2\over (x^2+1+b_1)^3},\nonumber\\
&
{dD_2\over dx}=({b_2D_2\over 12\pi^2}+{cD_3\over \pi^2}){x^2\over
(x^2+1+b_1)^2}-{b_2c^2\over 12\pi^2}{x^2\over (x^2+1+b_1)^3},\nonumber\\
&
{dD_3\over dx}=({b_2D_3\over 6\pi^2}+{3D_4c\over 2\pi^2})
{x^2\over (x^2+1+b_1)^2}-{b_2^2c\over 72\pi^2}{x^2\over (x^2+1+b_1)^3},
\nonumber\\
&
{dD_4\over dx}={b_2D_4\over 4\pi^2}{x^2\over (x^2+1+b_1)^2}-
{b_2^3\over 1296\pi^2}{x^2\over (x^2+1+b_1)^3}.
\label{rg6}
\end{eqnarray}

\vskip .5truecm
{\it 2. Flow-equations for the couplings of all dimension 6 non-derivative
operators of the
electroeak theory upon integration over $A_0$}
\vskip .3truecm
The extended Lagrangian density is parametrised as
\begin{eqnarray}
&
L={1\over 4}F_{ij}F_{ij}+{1\over 2}(D_i\phi)^{\dagger}(D_i\phi)+{1\over 2}
(D_iA_0)^2+{1\over 2}A_1(K)\phi^{\dagger}\phi+
{1\over 24}A_2(K)(\phi^{\dagger}\phi)^2
\nonumber\\
&
{}~~~~~~+{1\over 2}B_1(K)A_0^2+{1\over 2}B_2(K)(A_0^2)^2+
C(K)\phi^{\dagger}\phi A_0^2+D(K)(\phi^{\dagger}\phi)^3\nonumber\\
&
{}~~~~~~+E(K)(\phi^{\dagger}\phi)^2A_0^2+
F(K)(A_0^2)^2\phi^{\dagger}\phi+G(K)(A_0^2)^3.
\end{eqnarray}

The starting ($K=\Lambda$) value of the couplings of
the new (dimension 6) operators  has been determined in \cite{jako1}:
\begin{eqnarray}
&
D={\zeta (3)\over 1024\pi^4}({3g^6\over 16}-{3\lambda g^4\over 8}
-{\lambda^2g^2\over 4}+{5\lambda^3\over 27}),~~~
E=-{\zeta (3)g^2\over 1024\pi^4}({109g^4\over 16}+
{47\lambda g^2\over 6}+{5\lambda^2\over 9}),\nonumber\\
&
F=-{\zeta g^6\over 64\pi^4},~~G=0.
\end{eqnarray}
One notices that these couplings are ${\cal O}(g^6)$, (here the counting
$\lambda\sim g^2$ is the convenient choice). It is assumed that this
classification will not change to the end of the $A_0$-integration.
Even more, we assume, that to accuracy ${\cal O}(g^6)$ $G(k)=0$.
All this has to be checked for selfconsistency at the end.

With these
assumptions a single Hubbard-Stratonovich transformation is sufficient to
reach an extended Lagrangian, formally quadratic in $A_0$:
\begin{eqnarray}
&
L={1\over 4}F_{ij}F_{ij}+{1\over 2}(D_i\phi)^{\dagger}(D_i\phi)+{1\over 2}
(D_iA_0)^2+{1\over 2}A_1(K)\phi^{\dagger}\phi+
{1\over 24}A_2(K)(\phi^{\dagger}\phi)^2
\nonumber\\
&
{}~~~~~~~~~~~~~~~~+{1\over 2}[B_1(K)+2\sqrt{a}\chi+2C(K)\phi^{\dagger}\phi
+2E(K)(\phi^{\dagger}\phi )^2]A_0^2\nonumber\\
&
{}~~~~~~~~~~~~~~~~+D(K)(\phi^{\dagger}\phi)^3-{1\over 2}\chi^2
\end{eqnarray}
with
\begin{equation}
a=B_2+2F\phi^{\dagger}\phi.
\end{equation}

After integrating out the high frequency part of $A_0$ in constant $\chi ,
A_0,\phi$ background one finds the following gap equation for $\tilde\chi =
2\sqrt{a}\chi$:
\begin{eqnarray}
&
\tilde\chi=2aA_0^2+3aI_1(\Delta ),\nonumber\\
&
I_n(\Delta )={1\over 4\pi^2}\int_k^K {dp p^2\over (p^2+\Delta )^2 },
{}~~\Delta=B_1+\tilde\chi+2C\phi^{\dagger}\phi+2E(\phi^{\dagger}\phi )^2.
\end{eqnarray}
Repeating the arguments, already presented in the main text, we find
the following contribution to the potential energy density of the
reduced theory:
\begin{equation}
\Delta U={3\over 4\pi^2}\sum_{n=1}^{\infty}{(-1)^{n+1}\over n}
(2B_2A_0^2+4FA_0^2\phi^{\dagger}\phi+2C\phi^{\dagger}\phi
+2E(\phi^{\dagger}\phi )^2)^nI_n(B_1)
\end{equation}
For the flow-equations truncated at dimension 6 operators one truncates this
series at the third term. The following coupled set of equations is arrived at:
\begin{eqnarray}
&
{dA_1\over dk}=-{3C\over \pi^2}{k^2\over k^2+B_1},
{}~~{dB_1\over dk}=-{3B_2\over \pi^2}{k^2\over k^2+B_1}\nonumber\\
&
{dA_2\over dk}=-{18E\over \pi^2}{k^2\over k^2+B_1}+{36C^2\over \pi^2}
{k^2\over (k^2+B_1)^2},~~
{dB_2\over dk}={3B_2^2\over \pi^2}{k^2\over (k^2+B_1)^2},\nonumber\\
&
{dC\over dk}=-{3F\over \pi^2}{k^2\over k^2+B_1}+{3B_2C\over \pi^2}
{k^2\over (k^2+B_1)^2},\nonumber\\
&
{dD\over dk}=-{2C^3\over \pi^2}{k^2\over (k^2+B_1)^3}+{3CE\over \pi^2}
{k^2\over (k^2+B_1)^2},\nonumber\\
&
{dE\over dk}=-{6B_2C^2\over \pi^2}{k^2\over (k^2+B_1)^3}+({3B_2E\over \pi^2}
+{6CF\over \pi^2}){k^2\over (k^2+B_1)^2}\nonumber\\
&
{dF\over dk}=-{6B_2^2C\over \pi^2}{k^2\over (k^2+B_1)^3}+{6B_2F\over \pi^2}
{k^2\over (k^2+B_1)^2},\nonumber\\
&
{dG\over dk}=-{2B_2^3\over\pi^2}{k^2\over (k^2+B_1)^3}.
\end{eqnarray}

At this stage we use the knowledge of the order of magnitude of the different
couplings partly based on the solution of the set of equation truncated
at dimension 4 operators (see main text), partly (for the new operators)
by assumptions, whose consistency should be checked at the end of the
calculation:
\begin{eqnarray}
&
A_1\sim{\cal O}(g^2),~~B_1-m_D^2\sim{\cal O}(g^4),~~A_2\sim{\cal O}(g^2),
\nonumber\\
&B_2\sim{\cal O}(g^4),~~C\sim{\cal O}(g^2),~~D,E,F\sim{\cal O}(g^6).
\end{eqnarray}

One instantly recognizes, that the right hand side of the equation for $G$
is $\sim{\cal O}(g^{12})$, therefore the assumption made for it is
self-consistent. Similarly the right hand sides of the equations for $E$ and
$F$ also vanish to ${\cal O}(g^6)$, therefore they stay with their starting
${\cal O}(g^6)$ values. Consistently omitting on the right hand sides all terms
whose order of magnitude is smaller than ${\cal O}(g^6)$, one obtains the
following simplified set of equations:
\begin{eqnarray}
&
{dA_1\over dk}=-{3C\over \pi^2}{k^2\over k^2+B_1},
{}~~{dB_1\over dk}=-{3B_2\over \pi^2}{k^2\over k^2+B_1}\nonumber\\
&
{dA_2\over dk}=-{36E\over \pi^2}{k^2\over k^2+B_1}+{36C^2\over \pi^2}
{k^2\over (k^2+B_1)^2},~~
{dB_2\over dk}={3B_2^2\over \pi^2}{k^2\over (k^2+B_1)^2},\nonumber\\
&
{dC\over dk}=-{3F\over \pi^2}{k^2\over k^2+B_1}+{3B_2C\over \pi^2}
{k^2\over (k^2+B_1)^2},\nonumber\\
&
{dD\over dk}=-{2C^3\over \pi^2}{k^2\over (k^2+B_1)^3}
\end{eqnarray}
with $E,F$ being constants.

It is important to notice, that the equations for $B_1$ and $B_2$ are
the same as before, and the solution (\ref{solphi}) is accurate to
${\cal O}(g^6)$. Then, one easily casts the equation of $C(k)$
 into the form:
\begin{equation}
{d\over dk}({C\over B_2})=-{3F\over\pi^2}{k^2\over k^2+B_1(k)}
{1\over B_2(k)}.
\end{equation}
Its solution is simply
\begin{equation}
C(k)=Q_1B_2(k)+{3F\over\pi^2}B_2(k)\int_k^{\Lambda}{p^2\over p^2+B_1(p)}
{1\over B_2(p)}.
\end{equation}
The value of $Q_1$ is determined from the values taken at $\Lambda$, therefore
$Q_1$ equals to its value determined before.

Similarly, one transforms the equation of $A_1$ and $A_2$ into
\begin{eqnarray}
&
{dA_2\over dk}=12{dB_2\over dk}
(Q_1+{3FB_2\over \pi^2}
\int_k^{\Lambda}{dpp^2\over p^2+B_1(p)}{1\over B_2(p)})
^2-{18E\over \pi^2}
{k^2\over k^2+B_1},\nonumber\\
&
{dA_1\over dk}={dB_1\over dk}
(Q_1+{3F\over \pi^2}
\int_k^{\Lambda}{dpp^2\over p^2+B_1(p)}{1\over B_2(p)}).
\end{eqnarray}
The running of these three couplings is influenced by the
dimension 6 operators only additively as it is shown by the next
explicit formulae. These equalities are ${\cal O}(g^6)$ accurate
solutions of the above flow equations. They demonstrate explicitly the
stability of the ${\cal O}(g^4)$ solution appearing in (\ref{3dhiggscoupl}):
\begin{eqnarray}
&
C(0)-C(\Lambda )=Q_1(B_2(0)-B_2(\Lambda ))+{3FB_2(0)\over \pi^2}
\int_0^{\Lambda}{dp p^2\over p^2+B_1(p)}{1\over B_2(p)},\nonumber\\
&
A_1(0)-A_1(\Lambda )=Q_1(B_1(0)-B_1(\Lambda ))-{3FB_2(0)\over \pi^2}
\int_0^{\Lambda}{dp p^2\over p^2+B_1(p)}{B_1(0)-B_1(p)\over B_2(p)},\nonumber\\
&
A_2(0)-A_2(\Lambda )=12Q_1^2(B_2(0)-B_1(\Lambda ))-{18E\over \pi^2}
\int_0^{\Lambda}{dp p^2\over p^2+B_1(p)}.
\end{eqnarray}

{\bf Acknowledgement}

This research has been supported by the bilateral Franco-Hungarian Balaton
Research Agreement (Project No.67).



\begin{thebibliography}{12}

\bibitem{appelquist}
T. Appelquist and R. Pisarski, Phys. Rev. {\bf D23} (1981) 2305

\bibitem{landsman}
N.P. Landsman, Nucl. Phys. {\bf B322} (1989) 498

\bibitem{reisz}
T. Reisz, Z. f. Physik {\bf C53} (1992) 169

\bibitem{karka}
L. K\"arkk\"ainen, P. Lacock, B. Petersson and T.Reisz, Nucl. Phys. {\bf B395}
(1993) 733

\bibitem{braaten}
E. Braaten and A. Nieto, Effective Field Theory Approach to High Temperature
Thermodynamics, NUHEP-TH-95-2, hep-ph/9501375

\bibitem{karsch}
for a recent review of the reduction in lattice QCD, see F. Karsch,
Nucl. Phys. {\bf B34} (Proc. Suppl.) (1994) 63

\bibitem{nicoll}
N.F.Nicoll, T.S. Chang and H.E. Stanley, Phys. Lett. {\bf A57} (1976) 7

\bibitem{wegner}
F.Wegner and A.Houghton, Phys. Rev. {\bf A8} (1973) 401

\bibitem{hasenfratz}
A. Hasenfratz and P. Hasenfratz, Nucl. Phys. {\bf B270} (1986) 685

\bibitem{margaritis}
A. Margaritis, G. \'Odor and A. Patk\'os, Z. f. Physik {\bf C39} (1988) 109

\bibitem{tetradis}
N. Tetradis and C. Wetterich, Nucl. Phys. {\bf B383} (1992) 197

\bibitem{senben}
S.-B. Liao and  J. Polonyi, Ann. Phys. (N.Y.) {\bf 222} (1993) 122

\bibitem{morris}
T. Morris, Phys. Lett. {\bf B334} (1994) 335

\bibitem{buchmuller}
W. Buchm\"uller, Z. Fodor, T. Helbig and D. Walliser, Ann. Phys. (N.Y.)
{\bf 234} (1994) 260

\bibitem{zwirner}
J.R. Espinosa, M. Quir\'os and F. Zwirner, Phys. Lett. {\bf B314} (1993) 206

\bibitem{kajantie1}
K. Kajantie, K. Rummukainen and M. Shaposhnikov, Nucl. Phys. {\bf B407} (1993)
356

\bibitem{jako1}
A. Jakov\'ac, K. Kajantie and A. Patk\'os, Phys. Rev. {\bf D49} (1994) 6810

\bibitem{farakos1}
K. Farakos, K. Kajantie, K. Rummukainen and M. Shaposhnikov, Nucl. Phys.
{\bf B425} (1994) 67

\bibitem{ringwald}
A. Ringwald and C. Wetterich, Nuvc. Phys. {\bf B334} (1990) 506

\bibitem{rebhan}
For a careful review of magnetic screening, see: A. Rebhan,
Non-Abelian Debye screening in One-Loop Resummed Perturbation Theory,
DESY-94-132

\bibitem{jako2}
A. Jakov\'ac and A. Patk\'os, Phys. Lett. {\bf B334} (1994) 391

\bibitem{linde}
A. Linde, Rep. Prog. Phys. {\bf 42} (1979) 389

\bibitem{kirznitz}
A.D. Linde and D.A. Kirzhnitz, Phys. Lett. {\bf 72} (1972) 471;
Ann. Phys. (N.Y.) {\bf 101} (1976) 195

\bibitem{jako3}
A. Jakov\'ac, Reduction of the N-component scalar model at two loop level
hep-ph/9502113

\bibitem{pisarski}
R.B. Pisarski, Nucl. Phys. {\bf B309} (1988) 476
\end{thebibliography}
\end{document}